%% file: decaf_oakland.tex
\documentclass[conference,compsoc]{IEEEtran}
\ifCLASSOPTIONcompsoc
  \usepackage[nocompress]{cite}
\else
  \usepackage{cite}
\fi
\ifCLASSINFOpdf
\else
\fi
\usepackage{amsmath}      
\usepackage{tikz}
\usepackage{multirow}
\usepackage{float}
\usepackage{booktabs}     
\usepackage{hyperref}     
\usepackage[capitalise,noabbrev]{cleveref}
\usepackage{subcaption}
\usepackage{colortbl}
\usepackage[frozencache=true,cachedir=minted-cache]{minted}
\usepackage{tcolorbox}
\usepackage{pifont}       
\tcbuselibrary{minted,skins}
\usepackage{pgfplots}
\pgfplotsset{compat=1.18}

\usepackage{xspace}

\setminted{
  fontfamily=cmtt,
  fontsize=\small,
  breaksymbolleft=\textcolor{gray}{\tiny$\hookrightarrow$},
  breaksymbolright={},
  breaksymbolsep=0.4em, breaksymbolindentleft=0pt
}

\definecolor{mintedhighlight}{RGB}{255,250,205}

\definecolor{alexcolor}{RGB}{0,128,0}
\definecolor{edcolor}{RGB}{0,0,255}
\definecolor{osbertcolor}{RGB}{255,0,0}

\newcommand{\gray}{\cellcolor{gray!15}}

\newcommand{\Ghidra}{Ghidra\xspace}
\newcommand{\Exebench}{\textsc{ExeBench}\xspace}

\newcommand{\Idioms}{\textsc{Idioms}\xspace}
\newcommand{\Idiomsmodel}{\texttt{Idioms-\allowbreak Gemma-\allowbreak 7b}\xspace}

\newcommand{\hexrays}{Hex-Rays\xspace}
\newcommand{\hexraysdecompiler}{Hex-Rays Decompiler\xspace}
\newcommand{\llmfordecompile}{\textsc{LLM4Decompile}\xspace}
\newcommand{\llmfordecompilemodel}{\texttt{LLM4Decompile-\allowbreak Ref-\allowbreak 22b-\allowbreak v2}\xspace}

\newcommand{\llmfordecompilemodelsix}{\texttt{LLM4Decompile-\allowbreak Ref-6.7b-v1.6}\xspace}
\newcommand{\llmfordecompilemodelone}{\texttt{LLM4Decompile-\allowbreak Ref-1.3b-v1.6}\xspace}

\newcommand{\Qwenthree}{\textsc{Qwen3-32b}\xspace}

\newcommand{\Decaf}{\textsc{Decaf}\xspace}
\newcommand{\DecafLLM}{\texttt{Decaf-\allowbreak Gen-\allowbreak 22b}\xspace}
\newcommand{\DecafReranker}{\texttt{Decaf-\allowbreak ReRanker-\allowbreak 32b}\xspace}

\newcommand{\DecafLLMone}{\texttt{Decaf-\allowbreak Gen-\allowbreak 1.3b}\xspace}
\newcommand{\DecafLLMsix}{\texttt{Decaf-\allowbreak Gen-\allowbreak 6.7b}\xspace}

\newcommand{\alex}[1]{}
\newcommand{\ed}[1]{}
\newcommand{\es}[1]{}
\newcommand{\osbert}[1]{}

\newcommand{\etal}{et al.\xspace}

\begin{document}
%
\title{Decaf: Improving Neural Decompilation with Automatic Feedback and Search}

\author{\IEEEauthorblockN{Alexander Shypula}
\IEEEauthorblockA{University of Pennsylvania\\
Philadelphia, Pennsylvania 19104\\
Email: shypula@seas.upenn.edu}
\and
\IEEEauthorblockN{Osbert Bastani}
\IEEEauthorblockA{University of Pennsylvania\\
Philadelphia, Pennsylvania 19104\\
Email: obastani@seas.upenn.edu}
\and
\IEEEauthorblockN{Edward Schwartz}
\IEEEauthorblockA{Carnegie Mellon University\\
Pittsburgh, Pennsylvania 15213\\
Email: eschwartz@cert.org}}


%


\maketitle

\begin{center}\itshape Preprint. Under review.\end{center}

\input{sections/abstract}


%
\IEEEpeerreviewmaketitle

\thispagestyle{plain}
\pagestyle{plain}

\input{sections/intro}

\input{sections/methodology}

\input{sections/experiments}

\input{sections/results}

\input{sections/discussion}

\input{sections/background_related}

\input{sections/threats_and_conclusion}

\ifCLASSOPTIONcompsoc
  \section*{Acknowledgments}
\else
  \section*{Acknowledgment}
\fi




\bibliographystyle{IEEEtran}
%



\bibliography{decaf_oakland}


\end{document}

%% file: sections/abstract.tex
\begin{abstract}
Decompilers are useful tools used in reverse engineering to understand compiled source code. Reconstructing source code from compiled binaries is a challenging task, because high-level syntax, identifiers, and custom data types are generally lost as the compiler translates human-readable code to low-level machine code. Deterministic decompilers are useful tools for binary analysis, but can struggle to infer idiomatic syntax and identifier names. Generative AI models are a natural fit for reconstructing high-level syntax, identifiers, and types, but they can still suffer by hallucinating improper programming constructs and semantics. Instead of attempting to improve neural decompilers with more data and more training, we argue that compiler feedback can be used to dramatically improve the semantic correctness of neural decompiler outputs via search. Our system: \textsc{Decaf} (\textbf{DEC}ompilation with \textbf{A}utomated \textbf{F}eedback) raises the neural decompilation rate from 26.0\% on ExeBench to 83.9\% on the \textsc{Real} \texttt{-O2} split without sacrificing similarity to the original source code. We also find our automatic feedback methodology is highly effective for improving weaker neural decompilation models. 

\end{abstract}

%% file: sections/intro.tex
\section{Introduction}

Decompilation is the process of taking low-level assembly code from an executable and translating it back into a form that resembles the original high-level source code. Reverse engineers employ decompilers to assist with several security tasks, including analyzing malware, discovering, understanding and exploiting vulnerabilities, and fixing legacy software.  
Understanding compiled executables in general is a challenging task, because many of the design elements that make source-code understandable, such as variable, types, and identifier names, are ``lost'' as the program is lowered from an abstract source-code representation to an execution-focused assembly code representation.
Fortunately, decompiler research, which has been studied for over 30 years, has shown that many of these abstractions can be at least partially recovered by \emph{traditional decompilers} which employ sophisticated program analysis to recover information about the original program's variables, types, and functions.  Traditional decompilers include both academic efforts~\cite{dream,sailr,schwartz:2013} and industrial decompilers such as Ghidra and Hex-Rays, which are typically employed by reverse engineers. 

Although these abstractions make decompiled code more understandable than assembly code, traditional decompilers output code that is non-idiomatic and significantly \emph{more} difficult to understand than the original source code.  To demonstrate these limitations and motivate our solution, we present a simple C function in \cref{sub:original} as our working example, and its decompilation using the traditional decompiler Ghidra in \cref{sub:ghidra}.  It is easy to see that Ghidra did not recover function names, type names, variable names, or comments.  

To address these limitations, researchers have been studying how to apply neural
learning techniques to guess or predict many forms of missing information, such
as proposing identifier names~\cite{lacomis2019dire,chen2022dirty,varbert,gennm}
and recovering meaningful types~\cite{chen2022dirty,typeforge,tygr,resym}. More
recently, researchers have begun to train neural models to decompile entire
functions, either starting from assembly code~\cite{armengol2024slade,katz2018},
or learning how to transform the output of a traditional decompiler into the
original source code~\cite{tan2024llm4decompile, dramko2025idioms}. 

\input{figures/working_example_fig}
\input{figures/working_example_asm_fig}
These neural algorithms are an intuitive fit for decompilation, as training
pairs can be mined by pairing compiled source code with the original source code
that generated them. Learning algorithms will then encourage models to infer
reasonable identifier names, types, and idiomatic syntax from the context
provided. For example, in \cref{sub:llm4decompile}, we show the output of
LLM4Decompile~\cite{tan2024llm4decompile}, a recent neural decompiler based on
large language models (LLMs).  The decompiled code is easy to read, with
meaningful identifier names and idiomatic structure.


While these neural models can guess more idiomatic identifiers, types, and
source code, they offer no guarantees, and are prone to hallucinating
semantically incorrect decompiled code and can even fail to generate code that
compiles or executes without errors.  We can also see this in
\cref{sub:llm4decompile}, where the decompiled code is \emph{functionally
incorrect} because it omits the critical break condition that is present on line
9 of the original source.  This is the problem we are trying to solve in this
paper: how can we produce decompiled code that is both \emph{idiomatic} and
\emph{functionally correct}?


To this end, we are motivated by a simple observation: the first output of a neural
model is not always the best, and sampling multiple candidates increases the
chance that at least one is correct.  Recent neural decompilation work
~\cite{tan2024llm4decompile,dramko2025idioms} considers only a single candidate
per input, leaving potential improvements untapped.  As we show later in the
paper (\cref{sec:upper-limit,fig:decompilation-scale-samples}), sampling 32 candidates from a
neural decompiler yields a 88\% chance that at least one candidate is
functionally correct, compared to just 60\% when considering only a single
candidate. 

In this work, we show how to amplify neural decompilation models to produce code
that is both \emph{idiomatic} and \emph{functionally correct} by generating
multiple candidate decompilations and \emph{automatically} selecting the most
promising one.  The central challenge lies in the selection step, which reduces
to the undecidable problem of binary verification.  Fortunately, we propose
several practical approximations that work well in practice.  Our
best-performing is a \emph{neural reranker} that we trained to score each
candidate based on how closely its compiled output matches the original binary.
For example, our neural decompiler emitted the candidates in
\cref{sub:ours,sub:oursbad} (among others) which are functionally correct and
incorrect, respectively.  In \Cref{fig:reranker-example} we show how our neural reranker detected that the candidate in
\cref{sub:ours} was closer to the target function than the incorrect candidate
in \cref{sub:oursbad}, and yielded the correct decompilation to the reverse
engineer despite the fact that the bytewise distance between these two programs and the reference bytes were equivalent.

We name this approach \textsc{Decaf}: \textbf{Dec}ompilation with
\textbf{A}utomatic \textbf{F}eedback. Our approach is general and can be used to amplify any neural decompiler that produces diverse candidates. Using
Decaf, we substantially advance the state-of-the art in neural decompilation
across numerous metrics. On the \textsc{ExeBench} stripped \textsc{Real} \texttt{-O2} compiled split, we attain a 83.9\% functionally
correct solve rate and a 70.9\% exact byte-wise match significantly outperforming the strongest prior neural model,
\textsc{Idioms} (26.0\% and 19.1\% respectively) on this split. Reranking generally yields substantial improvements in functional correctness, and generally re-prioritizes decompiled programs that are more similar to the original source code. 


\noindent\textbf{Contributions}. In our work we contribute a new LLM generator model \DecafLLM that strongly outperforms other state-of-the-art neural decompilers on all \Exebench splits utilizing a single sample. We also contribute methodologies on how to benefit from taking more samples from a LLM decompiler, culminating with large LLM reranking model \DecafReranker which generally attains the highest reranking results on our evaluation sets. We demonstrate as well that our reranking methods also can also dramatically amplify the strength of other LLM decompilers such as \llmfordecompilemodel. We document potential weaknesses of our approach and outline potential directions for future work to benefit from our insights. We also open source our code and models to the community available at \url{https://github.com/AlexShypula/decaf}.

%% file: figures/working_example_fig.tex
\begin{figure*}[p]
  \centering

  \begin{subfigure}[t]{0.55\linewidth}
    \caption{Original Source Code}
    \label{sub:original}
    \begin{minted}[linenos, breaklines, breakanywhere, highlightlines={9}]{c}
void createDimensions(int* arr, int size){
  int x = 1, y;
  while(1){
    if(x == size){
      y = 1; break;
    }  
    y = size / x;
    if(x >= y){
      if(x * y == size) break;
    }
    x++;
  }
  arr[0] = x; arr[1] = y;
}
    \end{minted}
  \end{subfigure}\hfill
  \begin{subfigure}[t]{0.41\linewidth}
    \caption{LLM4Decompile Decompiled (Incorrect)}
    \label{sub:llm4decompile}
    \begin{minted}[linenos, breaklines, breakanywhere,
                  highlightlines={}, highlightcolor=mintedhighlight]{c}
void SplitBij(int *r, int n)
{
  int i, j;
  for (i = 1; i != n; i++) {
    j = n / i;
    if (i >= j) break;
  }
  r[0] = i; r[1] = n / i;
}
    \end{minted}
  \end{subfigure}

  \medskip

  \begin{subfigure}[t]{0.99\linewidth}
    \caption{Ghidra Decompiled (Correct)}
    \label{sub:ghidra}
    \begin{minted}[linenos, breaklines, breakanywhere, highlightlines={7}]{c}
void FUN_00101140(int *param_1,int param_2)
{
  int iVar1, iVar2 = 1;
  if (param_2 != 1) {
    do {
      iVar1 = param_2 / iVar2;
      if ((iVar1 <= iVar2) && (iVar1 * iVar2 == param_2)) goto LAB_0010116e;
      iVar2 = iVar2 + 1;
    } while (param_2 != iVar2);
  }
  iVar1 = 1;
LAB_0010116e:
  *param_1 = iVar2;
  param_1[1] = iVar1;
  return;
}
    \end{minted}
  \end{subfigure}

  \medskip

  \begin{subfigure}[t]{0.49\linewidth}
    \caption{Decaf Decompiled (Correct)}
    \label{sub:ours}
    \begin{minted}[linenos, breaklines, breakanywhere, highlightlines={9}]{c}
struct image_t {int cols; int rows; };
static void get_shape(image_t *self, int area)
{
  if (area == 1) {
    self->cols = 1; self->rows = 1; return;
  }

  int cols, rows = 1; cols = area / rows;
  while (!(cols <= rows && rows * cols == area))
  {
    rows++; cols = area / rows;
  }
  self->cols = rows; self->rows = cols;
}
    \end{minted}
  \end{subfigure}\hfill
  \begin{subfigure}[t]{0.45\linewidth}
    \caption{Decaf Decompiled (Incorrect)}
    \label{sub:oursbad}
    \begin{minted}[linenos, breaklines, breakanywhere,
                  highlightlines={}]{c}
struct Image {int width; int height; };
void get_image_size(Image *image, int number_of_pixels) {
  int i;

  for (i = 1; i < number_of_pixels; ++i) {
    if (i * i > number_of_pixels) {
      break;
    }
  }

  image->width = i;
  image->height = number_of_pixels / i;
}
    \end{minted}
  \end{subfigure}

  \caption{Working example. Code is slightly reformatted for space. (\subref{sub:original}) The original source code returns a factorization of \mintinline{c}{size}.  The highlighted line shows the critical break condition logic that is missing from some decompilations.
  (\subref{sub:llm4decompile}) LLM4Decompile is easy to understand but omits the critical break condition.
  (\subref{sub:ghidra}) Ghidra is functionally correct but non-idiomatic.
  (\subref{sub:ours}) Decaf's first choice is easy to understand and functionally correct.
  (\subref{sub:oursbad}) Lower ranked Decaf candidates are close but functionally incorrect, similar to LLM4Decompile.}
  \label{fig:working}
\end{figure*}

%% file: figures/working_example_asm_fig.tex

\definecolor{refbg}{RGB}{245,247,252}
\definecolor{correctbg}{RGB}{243,250,244}
\definecolor{incorrectbg}{RGB}{252,244,244}
\definecolor{refborder}{RGB}{120,130,170}
\definecolor{correctborder}{RGB}{80,160,90}
\definecolor{incorrectborder}{RGB}{180,75,75}
\definecolor{scoregoodbg}{RGB}{220,240,222}
\definecolor{scorebadbg}{RGB}{245,220,220}
\definecolor{hlref}{RGB}{200,215,245}
\definecolor{hlcorrect}{RGB}{195,230,200}
\definecolor{hlincorrect}{RGB}{245,195,195}
\definecolor{addr}{RGB}{120,120,130}

\newcommand{\A}[1]{\textcolor{addr}{#1}}

\newlength{\asmboxheight}
\setlength{\asmboxheight}{16em}

\begin{figure*}[h!]
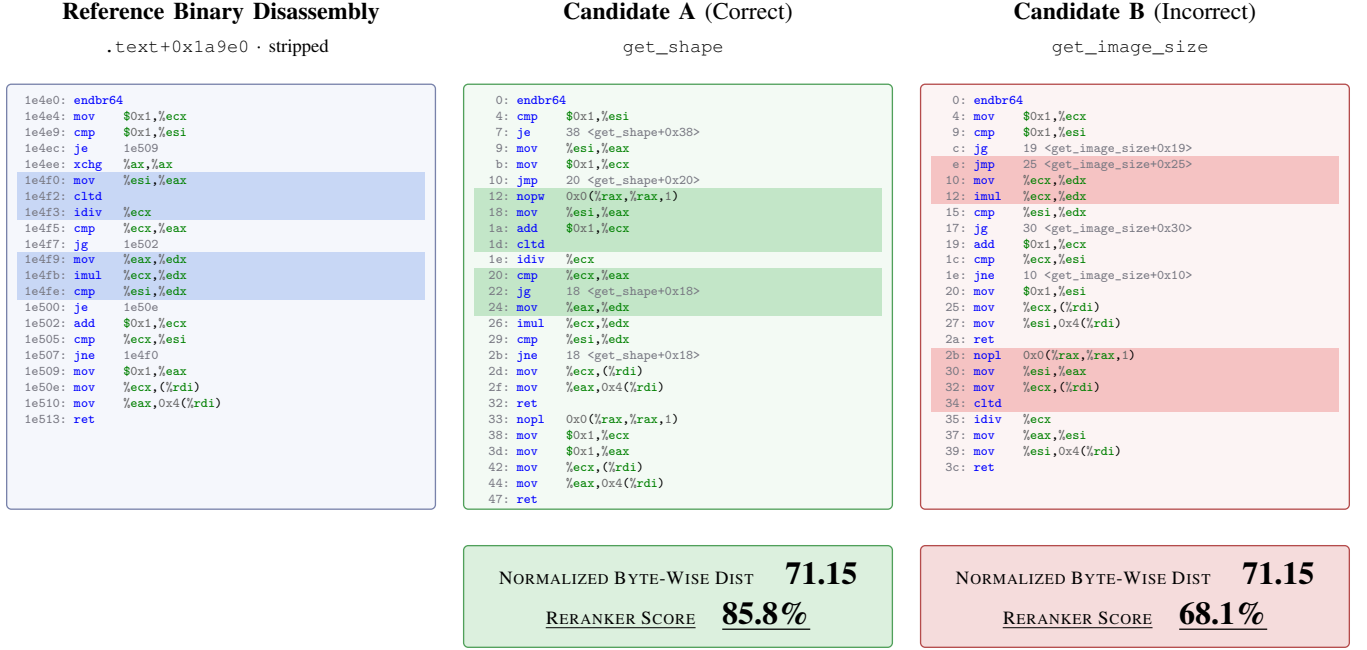

\centering

\begin{minipage}[t]{0.32\textwidth}
\centering{\small\textbf{Reference Binary Disassembly}}\\[1pt]
{\scriptsize\texttt{.text+0x1a9e0} $\cdot$ stripped}
\vspace{3pt}
\begin{tcolorbox}[
  colback=refbg, colframe=refborder,
  boxrule=0.5pt, arc=1.5pt,
  left=1pt, right=1pt, top=0pt, bottom=0pt,
  height=\asmboxheight,
  valign=top,
]
\begin{minted}[fontsize=\tiny, breaklines, escapeinside=||, highlightlines={6-8,11-13}, highlightcolor=hlref]{nasm}
 |\A{1e4e0:}| endbr64
 |\A{1e4e4:}| mov    $0x1,%ecx
 |\A{1e4e9:}| cmp    $0x1,%esi
 |\A{1e4ec:}| je     |\A{1e509}|
 |\A{1e4ee:}| xchg   %ax,%ax
 |\A{1e4f0:}| mov    %esi,%eax
 |\A{1e4f2:}| cltd
 |\A{1e4f3:}| idiv   %ecx
 |\A{1e4f5:}| cmp    %ecx,%eax
 |\A{1e4f7:}| jg     |\A{1e502}|
 |\A{1e4f9:}| mov    %eax,%edx
 |\A{1e4fb:}| imul   %ecx,%edx
 |\A{1e4fe:}| cmp    %esi,%edx
 |\A{1e500:}| je     |\A{1e50e}|
 |\A{1e502:}| add    $0x1,%ecx
 |\A{1e505:}| cmp    %ecx,%esi
 |\A{1e507:}| jne    |\A{1e4f0}|
 |\A{1e509:}| mov    $0x1,%eax
 |\A{1e50e:}| mov    %ecx,(%rdi)
 |\A{1e510:}| mov    %eax,0x4(%rdi)
 |\A{1e513:}| ret
\end{minted}
\end{tcolorbox}
\end{minipage}%
\hfill
\begin{minipage}[t]{0.32\textwidth}
\centering{\small\textbf{Candidate A} (Correct)}\\[1pt]
{\scriptsize\texttt{get\_shape}}
\vspace{3pt}
\begin{tcolorbox}[
  colback=correctbg, colframe=correctborder,
  boxrule=0.5pt, arc=1.5pt,
  left=1pt, right=1pt, top=0pt, bottom=0pt,
  height=\asmboxheight,
  valign=top,
]
\begin{minted}[fontsize=\tiny, breaklines, escapeinside=||, highlightlines={7-10,12-14}, highlightcolor=hlcorrect]{nasm}
   |\A{0:}| endbr64
   |\A{4:}| cmp    $0x1,%esi
   |\A{7:}| je     |\A{38 <get\_shape+0x38>}|
   |\A{9:}| mov    %esi,%eax
   |\A{b:}| mov    $0x1,%ecx
  |\A{10:}| jmp    |\A{20 <get\_shape+0x20>}|
  |\A{12:}| nopw   0x0(%rax,%rax,1)
  |\A{18:}| mov    %esi,%eax
  |\A{1a:}| add    $0x1,%ecx
  |\A{1d:}| cltd
  |\A{1e:}| idiv   %ecx
  |\A{20:}| cmp    %ecx,%eax
  |\A{22:}| jg     |\A{18 <get\_shape+0x18>}|
  |\A{24:}| mov    %eax,%edx
  |\A{26:}| imul   %ecx,%edx
  |\A{29:}| cmp    %esi,%edx
  |\A{2b:}| jne    |\A{18 <get\_shape+0x18>}|
  |\A{2d:}| mov    %ecx,(%rdi)
  |\A{2f:}| mov    %eax,0x4(%rdi)
  |\A{32:}| ret
  |\A{33:}| nopl   0x0(%rax,%rax,1)
  |\A{38:}| mov    $0x1,%ecx
  |\A{3d:}| mov    $0x1,%eax
  |\A{42:}| mov    %ecx,(%rdi)
  |\A{44:}| mov    %eax,0x4(%rdi)
  |\A{47:}| ret
\end{minted}
\end{tcolorbox}
\vspace{2pt}
\begin{tcolorbox}[
  colback=scoregoodbg, colframe=correctborder,
  boxrule=0.5pt, arc=1.5pt,
  left=2pt, right=2pt, top=3pt, bottom=3pt,
]
\centering

{\scriptsize\textsc{Normalized Byte-Wise Dist }}\quad{\large\textbf{71.15}} 
 \\[4pt]
{\scriptsize\underline{\textsc{Reranker Score}}}\quad{\large\underline{\textbf{85.8\%}}}

\end{tcolorbox}
\end{minipage}%
\hfill
\begin{minipage}[t]{0.32\textwidth}
\centering{\small\textbf{Candidate B} (Incorrect)}\\[1pt]
{\scriptsize\texttt{get\_image\_size}}
\vspace{3pt}
\begin{tcolorbox}[
  colback=incorrectbg, colframe=incorrectborder,
  boxrule=0.5pt, arc=1.5pt,
  left=1pt, right=1pt, top=0pt, bottom=0pt,
  height=\asmboxheight,
  valign=top,
]
\begin{minted}[fontsize=\tiny, breaklines, escapeinside=||, highlightlines={5-7,17-20}, highlightcolor=hlincorrect]{nasm}
   |\A{0:}| endbr64
   |\A{4:}| mov    $0x1,%ecx
   |\A{9:}| cmp    $0x1,%esi
   |\A{c:}| jg     |\A{19 <get\_image\_size+0x19>}|
   |\A{e:}| jmp    |\A{25 <get\_image\_size+0x25>}|
  |\A{10:}| mov    %ecx,%edx
  |\A{12:}| imul   %ecx,%edx
  |\A{15:}| cmp    %esi,%edx
  |\A{17:}| jg     |\A{30 <get\_image\_size+0x30>}|
  |\A{19:}| add    $0x1,%ecx
  |\A{1c:}| cmp    %ecx,%esi
  |\A{1e:}| jne    |\A{10 <get\_image\_size+0x10>}|
  |\A{20:}| mov    $0x1,%esi
  |\A{25:}| mov    %ecx,(%rdi)
  |\A{27:}| mov    %esi,0x4(%rdi)
  |\A{2a:}| ret
  |\A{2b:}| nopl   0x0(%rax,%rax,1)
  |\A{30:}| mov    %esi,%eax
  |\A{32:}| mov    %ecx,(%rdi)
  |\A{34:}| cltd
  |\A{35:}| idiv   %ecx
  |\A{37:}| mov    %eax,%esi
  |\A{39:}| mov    %esi,0x4(%rdi)
  |\A{3c:}| ret
\end{minted}
\end{tcolorbox}
\vspace{2pt}
\begin{tcolorbox}[
  colback=scorebadbg, colframe=incorrectborder,
  boxrule=0.5pt, arc=1.5pt,
  left=2pt, right=2pt, top=3pt, bottom=3pt,
]
\centering

{\scriptsize\textsc{Normalized Byte-Wise Dist }}\quad{\large\textbf{71.15}} 
 \\[4pt]

{\scriptsize\underline{\textsc{Reranker Score}}}\quad{\large\underline{\textbf{68.1\%}}}
\end{tcolorbox}
\end{minipage}

\vspace{6pt}

\caption{
A comparison of the disassembly used for reranking the examples from \Cref{fig:working}.
  \textbf{Candidate~A} preserves the per-iteration \texttt{idiv}/\texttt{imul}
  sequence found in the reference (85.8\%).
  \textbf{Candidate~B} lacks \texttt{idiv} in its loop body entirely
  (highlighted), instead deferring it to a separate exit path after
  \texttt{ret} at \texttt{0x30} (68.1\%)
}
\label{fig:reranker-example}
\end{figure*}

%% file: sections/methodology.tex
\section{Methodology}

\subsection{Approach}

\input{figures/decaf_schema_fig}

We provide an overview of our multi-step process in \Cref{fig:decaf_overview}. Following some prior work in neural decompilation we first process the target function with a \emph{traditional} decompiler~\cite{tan2024llm4decompile,dramko2025idioms}.
Traditional decompilers have been studied for decades~\cite{VanEmmerik2001}, and this allows neural decompilers to benefit from hard-earned engineering advances.  In this paper, we utilize the \Ghidra decompiler, because it is widely and freely available.

We provide \Ghidra's decompilation output as the \emph{input} to our neural decompiler model. Similar to other recent neural decompilers~\cite{tan2024llm4decompile,dramko2025idioms}, when given the output from a traditional decompiler, our model is trained to predict the original source code that was compiled into the target function.  (See~\cref{subsec:data}).

This is where our approach begins to differ from prior work.  
Rather than performing inference a single time, we can take an arbitrarily large number of samples from our model (e.g., 32, 128, 1024) using a standard ML technique called temperature sampling~\cite{ackley1985learning, ficler2017controlling, guo2017calibration}. The advantage of temperature sampling is that it introduces randomness into the inference process, which enables the model to produce a diverse set of responses.  As we will show later in this paper, the best decompilation is often not the first. 

At this point, we have sampled the model $n$ times and have $n$ candidate decompilations.  We attempt to re-compile each of these candidates into an object file (e.g., using \texttt{gcc -c}). If a candidate fails to compile, we discard it. We are now left with $m$ candidate decompilations that compile in isolation.  The next step is to identify the candidate that most closely resembles the original.  To do this, we employ a ``reranker'' neural model that takes two assembly sequences and predicts how likely they are functionally equivalent.  We use the reranker to compare each of the $m$ compiled candidates to the target assembly code, and return the top candidates to the user.

\subsection{Data}
\label{subsec:data}

\noindent\textbf{LLM Decompiler Data}. Obtaining training data was crucial for training both our LLM decompiler model as well as our LLM reranker model. To create training data for fine-tuning both LLM models, we used \Exebench \cite{armengol2022exebench}. \Exebench is a large scale benchmark of standalone compilable and executable C functions mined from open source code. 
We obtained source code from \Exebench, which we subsequently compiled and decompiled to create training pairs.  


We processed the \Exebench \textsc{Real Simple IO}, \textsc{Synth Rich IO}, and \textsc{Synth Simple IO} splits for training our LLM decompiler, and used the \textsc{Real Simple IO} and \textsc{Synth Rich IO} splits for our neural reranker. \Exebench provides heuristically inferred \texttt{\#include} statements for the \textsc{Real} split and leverages \textsc{Psyche-C} \cite{melo2020psychec, da2021anghabench}, a type inference tool for incomplete C programs, for the \textsc{Synth} splits. We incorporate both the inferred headers and type information into our prediction task, enabling \emph{type-aware supervision} where the model is trained to produce compilable code. Although \llmfordecompile \cite{tan2024llm4decompile} also required a large-scale dataset of \Ghidra-decompiled functions, only 100,000 examples were publicly released, and these were not processed with the additional steps we outline below. 



We compiled all functions using \texttt{gcc} 11.4.0 at both the \texttt{-O0} and \texttt{-O2} optimization levels. All programs were compiled for \texttt{x86-64} architecture. To simulate more real-world decompilation scenarios, we compile each \Exebench example with a dummy \texttt{main} function to produce a linked executable file instead of an object file. 

Other research has frequently found that decompiler output is challenging to recompile \cite{tan2024llm4decompile}.  We believe that this is partly caused by attempting to compile a function's decompiled code without declaring functions, types and globals the function may reference.  To mitigate this problem, we utilize a custom decompilation exporter for Ghidra that properly declares referenced functions, types, and globals using its internal databases. In \Cref{tab:results} we demonstrate that our custom exporter dramatically enhances the re-compilability and  executability of \Ghidra decompiled code. 
For each \Exebench function that we compile, we produce two decompilations: (1) from the unstripped compiled binary that contains symbols, and (2) from the stripped executable (i.e., processed with the \texttt{strip} command). 
The collection of all this data allows us to create a dataset where the traditionally decompiled input we can feed to our neural decompiler along with the original source code and dependencies that we will train to predict. 

\input{tables/generator_dataset_statistics}

\noindent\textbf{LLM Reranker Data}. To train our LLM reranker, we require a mixture of functionally equivalent \emph{and} non-equivalent decompiled functions relative to the target function. To approximate the distribution of positive and negative examples we expect to encounter in practice---a common best practice in ML---we construct such examples using a multi-step pipeline.  First, we attempt to decompile functions from the \emph{training} split using \Ghidra.  Next, we decompile the same functions using \llmfordecompilemodel, sampling eight candidate outputs per function. Finally, we compile and execute each  LLM-generated candidate to determine functional correctness. 

We found that often the outputs of executing reference functions from \Exebench did not match the cached outputs provided in the dataset itself. We re-processed our  executable splits by compiling and re-executing all possible functions twice (a second time to ensure programs did not contain side-effects from pseudo-random libraries or other sources of non-determinism). 

Then, using each of our LLM-generated outputs, we cache the output and side-effects from execution of all test cases. This enables comparison against the target function as well as comparison between generations to form positive and negative pairs based on execution-based equivalence. For each function, we can further expand the training examples by re-compiling the decompiled function at both the \texttt{-O0} and \texttt{-O2} optimization levels.

Because we evaluate on both stripped and unstripped binaries, we collect both forms of disassembly for these verification examples. In practice, especially for code compiled at \texttt{-O0}, disassembly can become quite long and may exceed the reranker's context window. Rather than including full disassembly pairs, we compress the input by pairing the target disassembly with a diff with respect to the generated disassembly.  All disassembly is obtained using  GNU binutils \texttt{objdump~-d}. 

\noindent\textbf{Vulnerability Recovery Experiment Data}. In addition to the \Exebench test set, we also evaluate \Decaf on its ability to generalize to other valid use cases such as recovering vulnerabilities in stripped binaries via decompilation. The premise of this experiment is to assess if decompilation can recover vulnerabilities in source code from stripped and compiled binaries which can then be detected by static analysis tools like CodeQL. 

For this we use version C/C++ Version 1.3 Juliet test suite\footnote{Available via: \url{https://github.com/arichardson/juliet-test-suite-c}}. The Juliet test suite is a part of the NIST Software Assurance Reference Dataset consisting of data sets with programs with documented weaknesses. The Juliet test suite contains C/C++ test cases organized into 118 different Common Weakness Enumeration (CWE) classes. Each test case contains a \emph{good} example without any vulnerability and a \emph{bad} example harboring a pattern within the CWE class. We subset only good/bad pairs which were classified correctly using CodeQL's C/C++ \texttt{security-extended} query suite, and then used random stratified sub-sampling of examples within each CWE to ensure high coverage and reduce experiment latency and re-balance the total number of good/bad functions to be even.  We report results on a function-level: where we process 296 total functions (148 each). Binaries are compiled from the Juliet test suite using the \texttt{-O2 -fno-inline} flags from \texttt{gcc} and are stripped of debug information. 


\input{tables/reranker_dataset_statistics}

\subsection{Implementation}

\textbf{Models.} For our LLM generation model, we utilize \llmfordecompilemodel as our base model. Our decision lies in the general intuition and empirical evidence that larger models trained on more data tend to perform better \cite{kaplan2020scaling}: the model consists of 22 billion parameters and was fine-tuned on a large corpus of decompilation data from \Exebench. To ablate decompilation performance relative to parameters, we also finetune \llmfordecompilemodelsix and \llmfordecompilemodelone on our dataset. For our LLM reranker model, we utilize \Qwenthree, a 32-billion parameter LLM \cite{yang2025qwen3}, given that models from a similar family appear to be among the best-performing open-source LLMs at equivalence checking for \texttt{x86-64} assembly \cite{wei2025equibench}.

\noindent\textbf{Training}. All training was performed on a DGX B200 node with 8$\times$ NVIDIA B200 GPUs (192GB HBM3e each), dual Intel Xeon Platinum 8570 processors (112 cores), and 2TB system memory using bfloat16 precision, FSDP, and gradient checkpointing. We trained two reranker models from \Qwenthree---one for stripped and one for unstripped binaries---for approximately 2 weeks each. The generator was fine-tuned from \llmfordecompilemodel for approximately 3 days (2 epochs). 

\noindent\textbf{Hyperparameters}. For the generator model, we used a learning rate of 2e-6 with 1000 warmup steps, an effective batch size of 528 (batch size 3 per GPU $\times$ 8 GPUs $\times$ 22 gradient accumulation steps), and a maximum sequence length of 5120 tokens. For the reranker models, we used a learning rate of 5e-6 with a linear scheduler and 15\% warmup ratio, an effective batch size of 256 (batch size 2 per GPU $\times$ 8 GPUs $\times$ 16 gradient accumulation steps), and a maximum sequence length of 4096 tokens. Both used the AdamW optimizer with gradient clipping (max norm 1.0).

\noindent\textbf{Inference and Evaluation}. Inference and evaluation were conducted on servers equipped with 8$\times$ NVIDIA RTX A6000 GPUs (48GB each), with either dual Intel Xeon Gold 6342 CPUs (48 cores, 96 threads) and 1TB RAM, or dual AMD EPYC 7402 CPUs (48 cores) and 504GB RAM. During data mining and evaluation, we executed hundreds of thousands of LLM-generated programs. Given the potential security risks of executing arbitrary generated code, all experiments were performed in a containerized environment. Our \texttt{Dockerfile} and scripts for installing additional dependencies are included in with our artifacts and will be made openly available.

\subsection{Evaluation Setup}
\label{subsec:metrics}

We evaluate decompilation quality along multiple axes:

\noindent\textbf{Functional correctness}. A decompilation is correct if, when recompiled, it produces a binary functionally equivalent to the original. In the general case, program verification is highly non-trivial, especially for programs with loops and in programming languages with pointers. As a result, we approximate functional correctness with test case execution. Using \Exebench, this entails re-compiling the decompiled code, linking it with a wrapper that instruments it with test cases, and checking that the outputs from executing the program as well as its side effects on the program state are equivalent to the reference function. We perform this for three of four dataset splits that are available to us. Because the \textsc{Synth} split may contain function calls and the stripping process will remove the names of these function calls: we do not have enough information to decompile and link these functions to the \Exebench harness containing implementations of these callee functions. As a result for the stripped \textsc{Synth} split we do not attempt to execute for correctness. Hence all fields for functional correctness for the striped \textsc{Synth} split remain empty given the challenge of linking a re-executing. 

\noindent\textbf{Byte-Wise Match}. In addition to measuring functional correctness through execution, we also impose a stricter notion of correctness by reporting byte-wise match of the re-compiled function. In our experiments, we assume we have access to the reference compiler. We consider this a relatively reasonable assumption given prior work has demonstrated that inferring the compiler which generated code is a task that can be predicted ~\cite{rosenblum2010compilerprovenance,du2023compilerprovenance}. Furthermore, it is a possibility that different compiler configurations could also be searched over as well. We define an alternate edit distance that accounts for relocation artifacts by first using binutils to identify the byte offsets affected by pending relocations. Bytes at these offsets are treated as wildcards, and we compute a modified Levenshtein distance that allows wildcard positions to match any character without penalty, focusing the distance on semantically meaningful differences. In our reranking experiments, we also use this modified edit distance as a method to rerank decompiled outputs. 

\noindent\textbf{Similarity to Reference Source Code}. Given that the goal of decompilation is often to assist the reverse engineer in understanding a program, a metric that approximates the similarity to the original source code is desirable. We use the Levenshtein distance, or the Edit distance, to approximate how similar our decompiled code is to the reference. The Edit distance metric we report is the Levenshtein distance normalized by the maximum length of the two sequences under consideration. 

\noindent\textbf{Compilation Rate}. In addition to all the metrics above, we also include the ability to compile the decompiled source code as an object file. For generations from \llmfordecompile, we allow the model to use \texttt{\#include} statements as well as other potential \textsc{Psyche-C} dependencies provided by \Exebench to enhance the compilability of code. 


\noindent\textbf{Models for Comparison and Modifications to Test Sets}. In order to report baseline comparisons, we use \llmfordecompilemodel from \cite{tan2024llm4decompile} and \Idiomsmodel from \cite{dramko2025idioms}. We use \Idiomsmodel as it has outperformed other models from \llmfordecompile on \Exebench and to our knowledge is state-of-the art on this task of any openly-available model. Because \Idiomsmodel requires inputs provided from the \hexraysdecompiler, we obtained \hexrays decompiled outputs for \Exebench in addition to \Ghidra decompiled outputs. Because over 15\% of some splits of \Exebench could not be decompiled by \hexraysdecompiler, we omitted these examples from the \Exebench test sets. In \Idioms, the authors focused on decompiling stripped inputs from \Exebench: as a result, we were only able to obtain stripped \hexrays inputs for the \Idiomsmodel and we could not evaluate \Idiomsmodel on the unstripped dataset splits. Because compiling and executing functions from \Exebench can be time consuming, we randomly subsampled 1,000 examples for all splits of the \Exebench test sets. For all experiments, when sampling from LLMs, we sample according to the distribution of the model with the Temperature parameter set to 0.8.

%% file: figures/decaf_schema_fig.tex
\begin{figure*}[t]
    \centering
    \includegraphics[width=\textwidth]{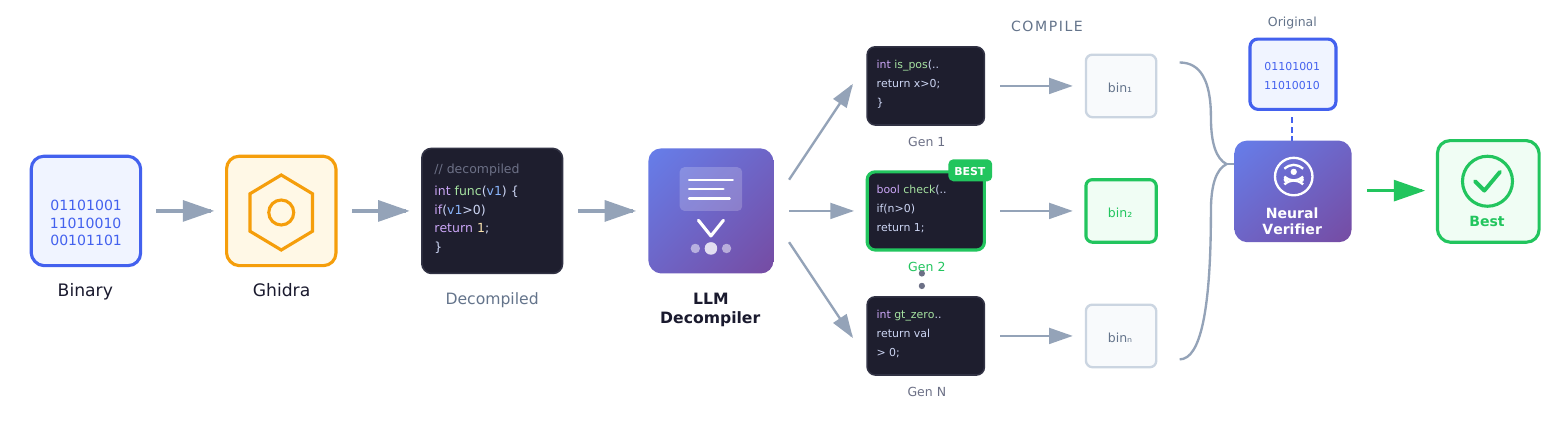}
    \caption{A Visual Overview of the \textsc{Decaf} system. In our implementation the \textsc{Decaf} pipeline follows multiple steps from taking the output from a traditional decompiler, generating multiple candidates from an LLM, compiling all the results, and finally getting feedback from our ``Verifier" LLM.}
    \label{fig:decaf_overview}
\end{figure*}

%% file: tables/generator_dataset_statistics.tex
\begin{table}[t]
\centering
\small
\caption{LLM fine-tuning dataset (1.34M unstripped / 1.35M stripped).}
\label{tab:llm-dataset}
\begin{tabular}{@{}lrrrr@{}}
\toprule
& \multicolumn{2}{c}{\textbf{Tokens}} & \multicolumn{2}{c}{\textbf{Lines}} \\
\cmidrule(lr){2-3} \cmidrule(lr){4-5}
& Prompt & Compl. & Prompt & Compl. \\
\midrule
\textbf{Unstripped} & & & & \\
\quad Total & 221.9M & 341.8M & 25.2M & 35.1M \\
\quad Mean & 166 & 255 & 19 & 26 \\
\quad Median & 98 & 178 & 13 & 21 \\
\midrule
\textbf{Stripped} & & & & \\
\quad Total & 244.9M & 343.9M & 24.3M & 35.3M \\
\quad Mean & 182 & 255 & 18 & 26 \\
\quad Median & 115 & 178 & 13 & 21 \\
\bottomrule
\end{tabular}
\end{table}

%% file: tables/reranker_dataset_statistics.tex
\begin{table}[t]
\centering
\small
\caption{Reranker dataset (1.52M unstripped / 1.80M stripped).}
\label{tab:reranker-dataset}
\begin{tabular}{@{}l r r@{}}
\toprule
& \textbf{Tokens} & \textbf{Lines} \\
\midrule
\textbf{Unstripped} & & \\
\quad Total & 1,982.8M & 181.5M \\
\quad Mean & 1,302 & 119 \\
\quad Median & 868 & 85 \\
\midrule
\textbf{Stripped} & & \\
\quad Total & 2,343.0M & 228.5M \\
\quad Mean & 1,299 & 127 \\
\quad Median & 859 & 90 \\
\bottomrule
\end{tabular}
\end{table}

%% file: sections/experiments.tex

%% file: sections/results.tex
\section{Experiments}
\input{tables/baselines_table}

We structure our evaluation around five research questions:

\textbf{RQ1:} How does model scale and fine-tuning with type-aware supervision affect decompilation performance?

\textbf{RQ2:} How does decompilation performance scale with more samples assuming perfect ``oracle" reranking. 

\textbf{RQ3:} In the absence of an execution oracle, how much performance can automatic feedback from compilation and neural reranking attain?

\textbf{RQ4:} Can automatic feedback from compilation be effective on other neural models? 

\textbf{RQ5:} Can \Decaf be effective in recovering vulnerabilities in compiled binaries? 

\textbf{RQ6:} Can neural reranking be robust to differences in compiler configuration 


\subsection{Model Scale and Type-Aware Supervision}

As previously mentioned, we adopt the underlying model \linebreak \llmfordecompilemodel; however, we further adapt it in two ways. To simulate more real-world reverse-engineering scenarios, we collected binaries that were stripped before decompiling so that we could adapt our model to this task. Additionally, we also trained our model to generate declarations for any functions, globals, or UDTs it may later reference within the source code. In this vein, our model is similar to \Idiomsmodel in that we jointly predict types and source code. However, our 22B model is significantly larger, trained on more decompilation examples and initialized from a strong pre-existing LLM. Our 6.7 and 1.3 billion parameter models are smaller than \Idiomsmodel. 

We report the performance of our own model as well as other baselines in \Cref{tab:results} using one sample only for evaluation. Across all splits our model \DecafLLM significantly outperforms \Idiomsmodel and \llmfordecompilemodel on functional accuracy and byte-wise match. For example on the stripped \textsc{Real} \texttt{-O2} split, our functional correctness performance is 59.1\%; whereas \llmfordecompilemodel and \Idiomsmodel are 24.0\% and 26.0\% respectively. Additionally we find that \DecafLLMsix and \DecafLLMone generally outperform \Idiomsmodel on functional accuracy and byte-wise match as well. We also find that \Ghidra with the custom decompilation exporter attains a very high compilability rate and functional accuracy rate; however, it underperfoms all neural models in terms of Edit-Distance relative to the source code. Additionally, we find that our model generally does not sacrifice much in terms of Edit-Distance relative to source code. Lastly, depending on the compiler flags used, we also find that even with one sample, our model can often achieve more byte-wise matches to the reference code than \Ghidra or the other LLM baselines: for example on the stripped \textsc{Real} \texttt{-O0} split \DecafLLM attains a 24.6\% byte-wise match rate compare to \Ghidra which attains a 17.2\% byte-wise match rate. 

\input{figures/scaling_real_stripped_O2_ours}

The only difference in our model and \llmfordecompilemodel is that our model [1] jointly predicts types, and [2] is also trained on stripped inputs. We believe both tasks may be contributing to performance improvements. When inspecting errors of \llmfordecompilemodel our model gets correct, the base model is prone to hallucinating UDTs in decompiled code that it does not generate declarations for. Nonetheless, the fact our model performs better on the stripped \textsc{Real} split than the unstripped \textsc{Real} is potential evidence that the LLM decompiler may be over-reliant on the function name as a shortcut heuristic instead of focusing on the semantics of the underlying source code. The phenomenon of shortcut learning in deep learning has been well-documented \cite{mccoy2019shortcut, geirhos2020shortcut}. Our model differs from \Idiomsmodel in that it is significantly larger, our dataset has more functions, and our input representation is \Ghidra. We hypothesize that the larger number of parameters and data our model has seen drive our performance.

\input{tables/reranking_table}

\subsection{Upper-Limit of Decompilation Performance with Samples} 
\label{sec:upper-limit}

We show in this section that taking more samples from a LLM significantly raises the ceiling for high-quality neural decompilations. In \Cref{fig:decompilation-scale-samples} we plot the best-possible performance across functional correctness, bytewise match, and source code edit distance as we take more samples. We report this on the \textsc{Real} split from \Exebench when compiled with \texttt{GCC -O2}. 

We find that as we take more samples, the upper-limit of performance can still dramatically increase: functional correctness best at 32 samples is 88.3\%, the lowest normalized edit distance is 39.7\%, and we can attain a 70.9\% byte-wise match rate. Assuming access to the reference compiler or an ability to search over compiler configurations, compiling the decompiled code and checking for byte-wise match is a highly-effective technique for searching over decompile examples. However, given no reference exact match may exist, improving the functional correctness of LLM-generated decompilations is still desirable.

\subsection{Improving Performance from Compilation and Neural Reranking}
\label{subsec:reranking_results}

\input{figures/reranking_scaling_real_stripped_O2}

In the scenario where we do not have access to an oracle for program execution, we attempt to approximate checking that our decompiled source code is equivalent to the original function we are trying to reverse. In \Cref{tab:reranking} we show our results from applying different methods of reranking to our task. 

In reranking, we use three methods. We adopt \textbf{Log Probability} reranking as a baseline where we use the length-normalized scoring function from \cite{wu2016google} to rerank candidate sequences. Since ranking by raw log-probability biases toward shorter sequences, the authors apply a length penalty to balance sequence quality against length:
\begin{equation}
\label{eq-rerank-scorer}
	\begin{split}
	s(Y, X) &= \log(P(Y | X)) / lp(Y) \\
	lp(Y) &= \frac{(5 + |Y|) ^ \alpha}{(5 + 1) ^ \alpha}
	\end{split}
\end{equation}
where $\log(P(Y | X)) = \sum_{t=1}^{|Y|} \log P(y_t | y_{<t}, X)$ is the sum of token-level log-probabilities from the generator model conditioned on the input prompt $X$, $|Y|$ is the length of the candidate sequence, and $\alpha$ controls the strength of the length normalization. We use $\alpha = 0.6$ in our experiments.

In addition to Log Probability reranking, we also investigate how utilizing the edit distance on the bytes between the reference function's bytes and the decompiled functions bytes (after re-compilation) as a reranking heuristic: we refer to this as our \textbf{Byte Dist} reranking method. This method relies on automatic compiler feedback, but does not require our neural reranker.

Lastly, we also evaluate our \textbf{neural reranker} \DecafReranker which takes as input the canonicalized disassembly of the reference function being decompiled and the disassembly of our LLM generated candidate function (after re-compilation). For this method, we employ a two-step process: if there is an exact byte-wise match after accounting for relocation alignment, we prioritize that function first. Otherwise, we take the generation that had the highest score under our neural reranker.

\noindent\textbf{Reranking Results}. Across the board, on nearly all splits, reranking with compiler feedback can substantially increase decompilation performance: we obtain byte-wise matches over 50\% of the time on all splits, and functional accuracy in the mid-eighty percents on numerous splits. For example, on the stripped \textsc{Real} \texttt{-O2} split, we attain a 83.9\% functionally
correct solve rate and a 70.9\% exact byte-wise match significantly outpacing our N=1 performance of 59.1\% and 38.0\% respectively 

Furthermore, even though reranking on the distance of compiled bytes or neural reranking on the functional correctness of assembly pairs does not directly optimize for similarity to source code: we see a strong pattern that reranking in this fashion generally improves the similarly to the source code: for example on the stripped \textsc{Real} \texttt{-O2} split our Edit Distance score drops from 59.2\% to 54.8\% with Byte Dist reranking and 55.5\% with neural reranking. In other words: reranking to improve functional correctness does not trade-off with similarity to the original source code: \emph{it actually improves it}. It is important to note that on the unstripped \texttt{-O2} \textsc{Synth} split, Log Probabilty reranking outperformed decompilation with compiler and neural feedback for functional correctness.

We note that at least within our experiments: there is a pattern that our neural reranker seems to consistently outperform Byte Distance reranking on functional accuracy; however, Byte Distance reranking outperforms our model in reranking decompiled functions that are more similar to the reference source functions, even in the stripped decompilation scenarios. Even though a compiler will substantially transform code, that bytes or assembly that are more similar to one-another are more likely to be generated by programs with similar procedural steps and programming syntax. As a result: at the extremes, our neural reranker which is optimized for correctness may ultimately tradeoff some idiomaticity in the implementation source code for a greater likelihood of generating a semantic correct decompilation.

\noindent\textbf{Scaling of Reranking Performance}. In \Cref{fig:reranking_scaling} we provide a more fine-grained analysis on how our reranking methods scale with more samples under consideration on the stripped \textsc{Real} split compiled with \texttt{-O2}. We note that although the general trend is that more samples generally improves performance, this does not imply that performance will necessarily monotonically improve. Especially with the neural reranker, there is a risk that as more programs are sampled, the potential may increase for a negative example to ``trick" the reranker into making a false-positive error. Despite some slight noise in the performance of the neural reranker over more samples considered, the fact the general trend is positive implies that likelihood of generating a positive example the reranker will correctly classify as positive is generally higher than the likelihood of generating a negative example that will ``trick" the reranker.

\subsection{Is Compilation and Neural Reranking Effective on Other Models?}
\label{sec:othermodels}

\input{tables/llm4decompile_rerank}

In addition to evaluating our reranking on our \Decaf generation model, we also evaluate it on \llmfordecompilemodel reranking over 32 generations in \Cref{tab:llmfordecomp_rerank}. 

We find that with a weaker decompilation model; our reranking methods can still yield very strong decompilation results: for example attaining 73.5\% functional correctness on decompiling \texttt{-O2} compiled and stripped functions from the \textsc{Real} split; significantly higher than the pass rate of 24.0\% attained by taking only one sample. Similar to the results discussed in \Cref{subsec:reranking_results}, we find that our neural reranker generally increases functional correctness over the Byte Dist reranking method with a tradeoff for a higher edit distance relative to the reference source code. These findings demonstrate that our method of using automatic compiler and neural feedback can be highly effective for other neural decompilation models.

\subsection{Can \Decaf be Effective in Recovering Vulnerabilities}

\input{tables/juliet_aggregate}

In addition to our experiments on decompiling for functional correctness on \Exebench, we also perform experiments on the NIST Juliet dataset as mentioned in \Cref{subsec:data}. We found greater success re-compiling and analyzing \Ghidra when we used the custom exporter mentioned in \Cref{subsec:data}. For the \Decaf experiments we use \DecafLLM with 32 samples. We report our results on function level analysis in \Cref{tab:juliet_table}. 

We see on this task that the Neural reranker achieves the highesst F1 score of all methods with 29.8\% and in general all \Decaf methods are capable of high-precision and reasonable recall: substantially outperforming Ghidra by over 20 points on F1. We hypothesize that this gap is in part because security-focused CodeQL queries may match against declarations of specific C standard library functions, which our model is trained to provide via forward declarations or \texttt{\#include} directives in its dependency block.

\subsection{Are Reranking Methods Robust to Alternative Compiler Configurations}
\label{subsec:stress_test}

In our work a major assumption made was access to the reference compiler configuration (e.g. either through a search process or through prediction). While prior work exists demonstrating that such configurations can often be predicted (see~\cref{sec:rel:provenance}), we stress test this idea by re-compiling our generated programs with  Clang instead of \texttt{GCC} for feedback. Specifically, for the splits on which we decompiled \texttt{GCC -O2} compiled code, we run two experiments where we assume we only have access to Clang with the \texttt{-O2} and \texttt{-Os} flags respectively. In \Cref{tab:clang_reranking} we show the results of this experiment: we omit byte-wise matching from the table as we were not able to find any bytewise matches after recompiling with Clang. 

We see two dramatically different stories between the stripped and unstripped splits. On the stripped splits, we see a similar trend as in all previous experiments where our neural reranker improves performance. While the neural reranker improves functional correctness, the overall functional accuracy and edit-distances are lower than before. For example on the stripped \textsc{Real} \texttt{-O2} split our neural reranker attains 70.5\% and 70.2\% functional correctness for code re-compiled with Clang \texttt{-O2} and \texttt{-Os} respectively: higher than N=1 performance of 59.1\%. However, on the unstripped split the phenomenon is very different. Log Probability reranking strongly outperforms on the \textsc{Real} split, and moreover neural reranking leads to degenerate performance on the \textsc{Synth} split. We hypothesize that our neural reranker generally struggles due to the distribution shift \cite{ben2010theory, yang2023rethinking}. While it may learn assembly semantics from its discriminative training task, it may not have been exposed to certain patterns of assembly the Clang may generate that \texttt{GCC} may not generate. This underscores an importance that a neural reranker should be trained on a highly diverse set of generated assembly pairs from different compiler configurations if it is expected to generalize to such scenarios. 

\input{tables/clang_compile_rerank}

%% file: tables/baselines_table.tex
\begin{table*}[h!]
\centering
\caption{Decompilation results across all splits with N=1 samples. \textbf{Acc}: functional correctness (percentage of decompilations that pass all I/O tests). \textbf{BM}: bytewise match (percentage of recompiled binaries that are byte-identical to the original). \textbf{Edit}: normalized source code edit distance between decompiled and original source (lower is better). \textbf{Comp}: compilation success rate. Ghidra + Cust.\ Exp.\ uses our custom exporter introduced in \Cref{subsec:data}. As discussed in \Cref{subsec:metrics} [1] we omit functional correctness on the stripped \textsc{Synth} split given the intractability of linking function calls after stripping the binary, and [2]  we only evaluate \Idiomsmodel on the stripped splits, because we only had access to stripped \hexrays inputs.}
\label{tab:results}
\resizebox{\textwidth}{!}{
\begin{tabular}{ll cccc cccc cccc cccc}
\toprule
& & \multicolumn{8}{c}{\textbf{Real}} & \multicolumn{8}{c}{\textbf{Synth}} \\
\cmidrule(lr){3-10} \cmidrule(lr){11-18}
& & \multicolumn{4}{c}{O0} & \multicolumn{4}{c}{O2} & \multicolumn{4}{c}{O0} & \multicolumn{4}{c}{O2} \\
\cmidrule(lr){3-6} \cmidrule(lr){7-10} \cmidrule(lr){11-14} \cmidrule(lr){15-18}
Split & Method & Acc & BM & Edit & Comp & Acc & BM & Edit & Comp & Acc & BM & Edit & Comp & Acc & BM & Edit & Comp \\
\midrule
\multirow{7}{*}{Stripped}
& Ghidra & 29.4 & 5.8 & 69.5 & 35.2 & 34.2 & 17.9 & 70.7 & 39.5 & -- & 0.8 & 76.3 & 8.5 & -- & 2.0 & 78.5 & 9.1  \\
& LLM4Decompile & 30.2 & 9.0 & 57.9 & 45.5 & 24.0 & 14.1 & \underline{61.2} & 43.1 & -- & 0.1 & 74.9 & 16.2 & -- & 0.3 & 77.8 & 16.1  \\
& Idioms Gemma & 42.8 & 16.5 & \textbf{54.9} & 69.3 & 26.0 & 19.1 & 62.2 & 60.5 & -- & 2.1 & \textbf{53.3} & 10.6 & -- & 2.4 & \textbf{58.5} & 11.2  \\
& \gray Ghidra + Cust.\ Exp. & \gray \textbf{83.1} & \gray 17.2 & \gray 69.5 & \gray \underline{95.2} & \gray \textbf{80.3} & \gray \textbf{40.8} & \gray 70.7 & \gray \underline{92.8} & \gray -- & \gray 9.2 & \gray 76.3 & \gray 81.2 & \gray -- & \gray 14.8 & \gray 78.5 & \gray 63.7  \\
& \gray \Decaf-1.3B & \gray 47.6 & \gray 17.8 & \gray 59.2 & \gray 89.7 & \gray 37.8 & \gray 25.4 & \gray 62.9 & \gray 87.1 & \gray -- & \gray 18.0 & \gray 64.8 & \gray 79.7 & \gray -- & \gray 18.7 & \gray 68.4 & \gray 79.7  \\
& \gray \Decaf-6.7B & \gray 54.9 & \gray \underline{21.0} & \gray 57.7 & \gray 91.3 & \gray 35.8 & \gray 29.3 & \gray \underline{61.2} & \gray 91.1 & \gray -- & \gray \underline{20.5} & \gray 62.8 & \gray \underline{87.4} & \gray -- & \gray \underline{21.4} & \gray 66.8 & \gray \underline{87.7}  \\
& \gray \Decaf-22B & \gray \underline{68.1} & \gray \textbf{24.6} & \gray \underline{56.0} & \gray \textbf{95.6} & \gray \underline{59.1} & \gray \underline{38.0} & \gray \textbf{59.2} & \gray \textbf{95.4} & \gray -- & \gray \textbf{28.6} & \gray \underline{59.1} & \gray \textbf{89.6} & \gray -- & \gray \textbf{28.2} & \gray \underline{62.9} & \gray \textbf{88.9}  \\
\midrule
\multirow{6}{*}{Unstripped}
& Ghidra & 29.3 & 5.8 & 64.4 & 34.8 & 33.2 & 17.1 & 65.7 & 38.4 & 7.0 & 0.8 & 64.4 & 8.0 & 6.5 & 2.0 & 70.5 & 9.1  \\
& LLM4Decompile & 45.9 & 18.1 & \textbf{44.5} & 60.6 & 38.8 & 24.7 & \textbf{46.9} & 56.9 & 25.6 & 16.2 & \underline{35.5} & 30.5 & 22.0 & 12.0 & 50.0 & 36.0  \\
& \gray Ghidra + Cust.\ Exp. & \gray \textbf{81.5} & \gray 17.2 & \gray 64.4 & \gray \underline{92.8} & \gray \textbf{79.3} & \gray \textbf{40.0} & \gray 65.7 & \gray 90.8 & \gray \textbf{75.1} & \gray 11.8 & \gray 64.4 & \gray \underline{85.6} & \gray \textbf{58.6} & \gray 20.4 & \gray 70.5 & \gray 81.5  \\
& \gray \Decaf-1.3B & \gray 43.0 & \gray 15.8 & \gray 51.1 & \gray 86.0 & \gray 38.0 & \gray 24.8 & \gray 53.7 & \gray 87.5 & \gray 33.6 & \gray 19.5 & \gray 46.9 & \gray 70.3 & \gray 31.7 & \gray 18.3 & \gray 52.3 & \gray 79.8  \\
& \gray \Decaf-6.7B & \gray 51.8 & \gray \underline{19.4} & \gray 48.8 & \gray 91.2 & \gray 44.4 & \gray 26.7 & \gray 52.1 & \gray \underline{92.3} & \gray 53.8 & \gray \underline{26.6} & \gray 37.5 & \gray 84.7 & \gray 37.2 & \gray \underline{22.0} & \gray \underline{48.2} & \gray \underline{87.9}  \\
& \gray \Decaf-22B & \gray \underline{61.3} & \gray \textbf{25.5} & \gray \underline{45.9} & \gray \textbf{94.9} & \gray \underline{53.8} & \gray \underline{34.9} & \gray \underline{49.5} & \gray \textbf{94.7} & \gray \underline{70.3} & \gray \textbf{40.9} & \gray \textbf{29.1} & \gray \textbf{90.4} & \gray \underline{47.5} & \gray \textbf{32.1} & \gray \textbf{40.6} & \gray \textbf{90.1}  \\
\bottomrule
\end{tabular}
}
\end{table*}

%% file: figures/scaling_real_stripped_O2_ours.tex
\begin{figure}[h]
\centering
\begin{tikzpicture}
\begin{axis}[
    title={Decompilation Upper Limit Scaling with Samples},
    title style={font=\bfseries, yshift=5pt},
    width=0.45\textwidth,
    height=0.4\textwidth,
    xlabel={Number of Samples ($k$)},
    ylabel={Success Rate},
    ylabel style={yshift=-5pt},
    xmin=0, xmax=34,
    ymin=0, ymax=1,
    xtick distance=8,
    ytick={0, 0.2, 0.4, 0.6, 0.8, 1.0},
    grid=both,
    grid style={line width=0.2pt, draw=gray!30},
    major grid style={line width=0.4pt, draw=gray!50},
    axis y line*=left,
    axis x line*=bottom,
]
\addplot[
    color=blue!70!black,
    line width=1.5pt,
] coordinates {(1,0.5910) (2,0.7000) (3,0.7540) (4,0.7750) (5,0.7930) (6,0.8060) (7,0.8120) (8,0.8240) (9,0.8310) (10,0.8370) (11,0.8410) (12,0.8500) (13,0.8530) (14,0.8580) (15,0.8590) (16,0.8600) (17,0.8630) (18,0.8640) (19,0.8640) (20,0.8660) (21,0.8660) (22,0.8660) (23,0.8690) (24,0.8700) (25,0.8730) (26,0.8750) (27,0.8760) (28,0.8780) (29,0.8790) (30,0.8810) (31,0.8830) (32,0.8830)};
\addplot[
    color=purple!70!black,
    line width=1.5pt,
    dashed,
] coordinates {(1,0.3800) (2,0.4750) (3,0.5350) (4,0.5680) (5,0.5860) (6,0.6030) (7,0.6140) (8,0.6230) (9,0.6300) (10,0.6350) (11,0.6440) (12,0.6520) (13,0.6580) (14,0.6630) (15,0.6670) (16,0.6710) (17,0.6740) (18,0.6760) (19,0.6800) (20,0.6840) (21,0.6850) (22,0.6870) (23,0.6890) (24,0.6940) (25,0.6960) (26,0.6970) (27,0.7000) (28,0.7000) (29,0.7040) (30,0.7070) (31,0.7080) (32,0.7090)};
\end{axis}
\begin{axis}[
    width=0.45\textwidth,
    height=0.4\textwidth,
    xmin=0, xmax=34,
    ymin=0, ymax=1,
    xtick=\empty,
    ytick={0, 0.2, 0.4, 0.6, 0.8, 1.0},
    ylabel={Edit Distance (lower is better)},
    ylabel style={yshift=5pt, orange!80!black},
    yticklabel style={orange!80!black},
    axis y line*=right,
    axis x line=none,
]
\addplot[
    color=orange!80!black,
    line width=1.5pt,
    dotted,
] coordinates {(1,0.5923) (2,0.5450) (3,0.5196) (4,0.5002) (5,0.4849) (6,0.4761) (7,0.4669) (8,0.4591) (9,0.4541) (10,0.4487) (11,0.4443) (12,0.4400) (13,0.4362) (14,0.4331) (15,0.4295) (16,0.4262) (17,0.4230) (18,0.4200) (19,0.4175) (20,0.4164) (21,0.4149) (22,0.4130) (23,0.4105) (24,0.4094) (25,0.4075) (26,0.4061) (27,0.4051) (28,0.4038) (29,0.4014) (30,0.3996) (31,0.3987) (32,0.3974)};
\end{axis}
\node[anchor=south east, draw=gray!50, fill=white, fill opacity=0.9, 
      font=\small, inner sep=5pt, align=left] at (rel axis cs:0.97,0.03) {
    \raisebox{2pt}{\tikz\draw[blue!70!black, line width=1.5pt] (0,0)--(0.5,0);} Functional Correctness \\[2pt]
    \raisebox{2pt}{\tikz\draw[purple!70!black, line width=1.5pt, dashed] (0,0)--(0.5,0);} Exact Bytewise Match \\[2pt]
    \raisebox{2pt}{\tikz\draw[orange!80!black, line width=1.5pt, dotted] (0,0)--(0.5,0);} Source Edit Distance
};
\end{tikzpicture}
\caption{We plot the best possible success rate for functional correctness, recompilation exact bytewise, and source code edit distance vs. number of samples taken. We plot this for our LLM generator model on the \textsc{Real} split from \Exebench when decompiling examples compiled with \texttt{GCC}'s \texttt{-O2} optimization flag.}
\label{fig:decompilation-scale-samples}
\end{figure}
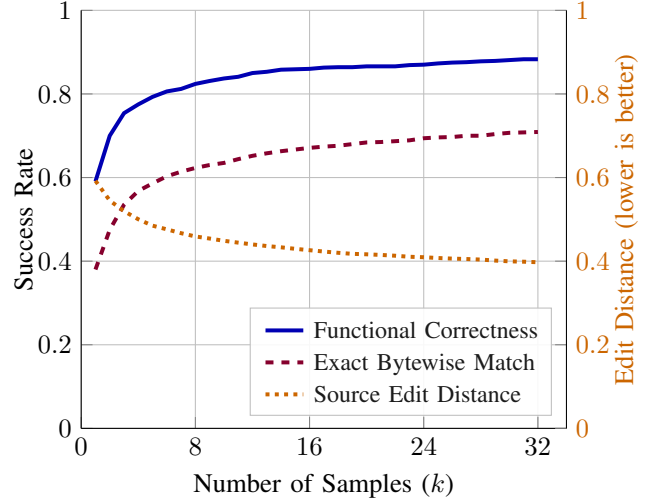

%% file: tables/reranking_table.tex
\begin{table*}[h!]
\centering
\caption{Reranking results with N=32 candidate samples using our \Decaf LLM Generator. \textbf{None}: single-sample baseline (N=1). \textbf{Log Prob}: reranking by model log probability. \textbf{Byte Dist}: reranking by bytewise distance to original binary. \textbf{\Decaf ReRanker}: learned reranking model. Metrics as in Table~\ref{tab:results}.}
\label{tab:reranking}
\resizebox{\textwidth}{!}{
\begin{tabular}{ll cccc cccc cccc cccc}
\toprule
& & \multicolumn{8}{c}{\textbf{Real}} & \multicolumn{8}{c}{\textbf{Synth}} \\
\cmidrule(lr){3-10} \cmidrule(lr){11-18}
& & \multicolumn{4}{c}{O0} & \multicolumn{4}{c}{O2} & \multicolumn{4}{c}{O0} & \multicolumn{4}{c}{O2} \\
\cmidrule(lr){3-6} \cmidrule(lr){7-10} \cmidrule(lr){11-14} \cmidrule(lr){15-18}
Split & Method & Acc & BM & Edit & Comp & Acc & BM & Edit & Comp & Acc & BM & Edit & Comp & Acc & BM & Edit & Comp \\
\midrule
\multirow{4}{*}{Stripped}
& None (N=1)           & 68.1 & 24.6 & 56.0 & 95.6 & 59.1 & 38.0 & 59.2 & 95.4 & --   & 28.6 & 59.1 & 89.6 & --   & 28.2 & 62.9 & 88.9 \\
& Log Prob             & 71.5 & 24.9 & 55.7 & 95.8 & 64.2 & 40.3 & 58.3 & 96.0 & --   & 31.4 & \underline{56.3} & \underline{90.5} & --   & 32.8 & \underline{60.1} & \underline{91.3} \\
& Byte Dist            & \underline{82.6} & \textbf{56.8} & \textbf{49.6} & \underline{99.3} & \underline{81.1} & \textbf{70.9} & \textbf{54.8} & \underline{99.4} & --   & \textbf{51.8} & \textbf{54.1} & \textbf{99.7} & --   & \textbf{46.7} & \textbf{58.5} & \textbf{99.6} \\
& \Decaf ReRanker      & \textbf{87.3} & \textbf{56.8} & \underline{51.3} & \textbf{99.7} & \textbf{83.9} & \textbf{70.9} & \underline{55.5} & \textbf{99.8} & --   & \textbf{51.8} & 55.3 & \textbf{99.8} & --   & \textbf{46.7} & 60.0 & \textbf{99.8} \\
\midrule
\multirow{4}{*}{Unstripped}
& None (N=1)           & 61.3 & 25.5 & 45.9 & 94.9 & 53.8 & 34.9 & 49.5 & 94.7 & 70.3 & 40.9 & 29.1 & 90.4 & 47.5 & 32.1 & 40.6 & 90.1 \\
& Log Prob             & 71.2 & 30.5 & 42.6 & 97.2 & 66.1 & 44.4 & 45.6 & 96.8 & 79.5 & 49.2 & \underline{23.7} & 93.3 & \textbf{62.5} & 38.9 & \underline{36.1} & \underline{95.0} \\
& Byte Dist            & \underline{83.5} & \textbf{59.2} & \textbf{38.5} & \underline{99.6} & \underline{82.0} & \textbf{71.1} & \textbf{44.2} & \underline{99.5} & \underline{84.6} & \textbf{62.2} & \textbf{22.9} & \underline{99.2} & 53.6 & \textbf{50.9} & \textbf{34.4} & \textbf{99.4} \\
& \Decaf ReRanker      & \textbf{86.5} & \textbf{59.2} & \underline{40.4} & \textbf{99.8} & \textbf{82.7} & \textbf{71.1} & \underline{44.9} & \textbf{99.7} & \textbf{86.9} & \textbf{62.2} & 25.0 & \textbf{99.3} & \underline{54.5} & \textbf{50.9} & 36.9 & \textbf{99.4} \\
\bottomrule
\end{tabular}
}
\end{table*}

%% file: figures/reranking_scaling_real_stripped_O2.tex
\begin{figure}[h!]
\centering
\begin{tikzpicture}
\begin{axis}[
    title={Reranking Method Scaling with Samples},
    title style={font=\bfseries, yshift=5pt},
    width=0.45\textwidth,
    height=0.4\textwidth,
    xlabel={Number of Samples ($k$)},
    ylabel={Functional Correctness},
    ylabel style={yshift=-5pt},
    xmin=0, xmax=34,
    ymin=0.55, ymax=0.95,
    xtick distance=8,
    ytick={0.55, 0.65, 0.75, 0.85, 0.95},
    grid=both,
    grid style={line width=0.2pt, draw=gray!30},
    major grid style={line width=0.4pt, draw=gray!50},
    axis y line*=left,
    axis x line*=bottom,
    legend style={
        at={(0.5,-0.22)},
        anchor=north,
        legend columns=2,
        font=\small,
        cells={anchor=west},
        draw=gray!50,
        fill=white,
        column sep=1em,
    },
]
\addplot[
    color=blue!70!black,
    line width=1.5pt,
] coordinates {(1,0.5910) (2,0.7000) (3,0.7540) (4,0.7750) (5,0.7930) (6,0.8060) (7,0.8120) (8,0.8240) (9,0.8310) (10,0.8370) (11,0.8410) (12,0.8500) (13,0.8530) (14,0.8580) (15,0.8590) (16,0.8600) (17,0.8630) (18,0.8640) (19,0.8640) (20,0.8660) (21,0.8660) (22,0.8660) (23,0.8690) (24,0.8700) (25,0.8730) (26,0.8750) (27,0.8760) (28,0.8780) (29,0.8790) (30,0.8810) (31,0.8830) (32,0.8830)};
\addlegendentry{Upper-Limit}
\addplot[
    color=purple!70!black,
    line width=1.5pt,
    dashed,
] coordinates {(1,0.5910) (2,0.6820) (3,0.7290) (4,0.7470) (5,0.7540) (6,0.7600) (7,0.7660) (8,0.7750) (9,0.7810) (10,0.7850) (11,0.7880) (12,0.7960) (13,0.8010) (14,0.8030) (15,0.8040) (16,0.8050) (17,0.8140) (18,0.8130) (19,0.8120) (20,0.8090) (21,0.8080) (22,0.8060) (23,0.8080) (24,0.8130) (25,0.8150) (26,0.8190) (27,0.8200) (28,0.8210) (29,0.8230) (30,0.8270) (31,0.8300) (32,0.8300)};
\addlegendentry{Neural Reranker}
\addplot[
    color=orange!80!black,
    line width=1.5pt,
    dotted,
] coordinates {(1,0.5910) (2,0.6730) (3,0.7160) (4,0.7300) (5,0.7440) (6,0.7530) (7,0.7550) (8,0.7580) (9,0.7620) (10,0.7700) (11,0.7740) (12,0.7830) (13,0.7860) (14,0.7910) (15,0.7920) (16,0.7950) (17,0.7990) (18,0.7980) (19,0.8000) (20,0.8000) (21,0.7990) (22,0.7990) (23,0.8030) (24,0.8050) (25,0.8060) (26,0.8090) (27,0.8100) (28,0.8090) (29,0.8110) (30,0.8120) (31,0.8120) (32,0.8110)};
\addlegendentry{Byte Distance}
\addplot[
    color=green!50!black,
    line width=1.5pt,
    dashdotted,
] coordinates {(1,0.5910) (2,0.6120) (3,0.6220) (4,0.6180) (5,0.6270) (6,0.6310) (7,0.6310) (8,0.6310) (9,0.6340) (10,0.6300) (11,0.6270) (12,0.6250) (13,0.6200) (14,0.6250) (15,0.6290) (16,0.6340) (17,0.6340) (18,0.6370) (19,0.6390) (20,0.6370) (21,0.6410) (22,0.6400) (23,0.6390) (24,0.6360) (25,0.6330) (26,0.6340) (27,0.6350) (28,0.6380) (29,0.6390) (30,0.6390) (31,0.6420) (32,0.6420)};
\addlegendentry{Log Probability}
\end{axis}
\end{tikzpicture}
\caption{Comparison of reranking methods for selecting among $n$ decompilation candidates. The upper-limit from executing all candidates and selecting the best. All reranking methods are also plotted where we vary the number of samples allowed to be reranked and choosing only the highest reranked example. We plot this on the \textsc{Real} split from \Exebench when decompiling examples compiled with \texttt{GCC}'s \texttt{-O2} optimization flag.}
\label{fig:reranking_scaling}
\end{figure}
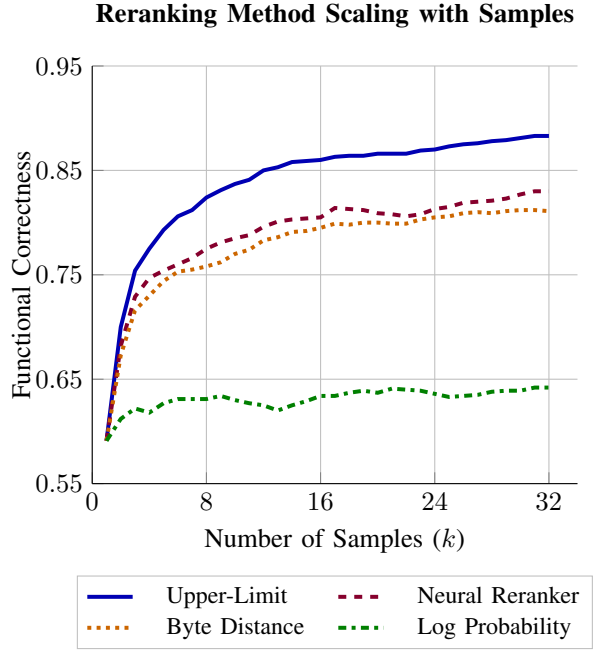

%% file: tables/llm4decompile_rerank.tex
\begin{table*}[h!]
\centering
\caption{Reranking results with N=32 candidate samples using LLM4Decompile. \textbf{None}: single-sample baseline (N=1). \textbf{Log Prob}: reranking by model log probability. \textbf{Byte Dist}: reranking by bytewise distance to original binary. Metrics: Acc (Accuracy \%), BM (Exact Match \%), Edit (C-Dist), Comp (Compile \%).}
\label{tab:llmfordecomp_rerank}
\resizebox{\textwidth}{!}{
\begin{tabular}{ll cccc cccc cccc cccc}
\toprule
& & \multicolumn{8}{c}{\textbf{Real}} & \multicolumn{8}{c}{\textbf{Synth}} \\
\cmidrule(lr){3-10} \cmidrule(lr){11-18}
& & \multicolumn{4}{c}{O0} & \multicolumn{4}{c}{O2} & \multicolumn{4}{c}{O0} & \multicolumn{4}{c}{O2} \\
\cmidrule(lr){3-6} \cmidrule(lr){7-10} \cmidrule(lr){11-14} \cmidrule(lr){15-18}
Split & Method & Acc & BM & Edit & Comp & Acc & BM & Edit & Comp & Acc & BM & Edit & Comp & Acc & BM & Edit & Comp \\
\midrule
\multirow{4}{*}{Stripped}
& None (N=1) & 30.2 & 9.0 & 57.9 & 45.5 & 24.0 & 14.1 & 61.2 & 43.1 & -- & 0.1 & 74.9 & 16.2 & -- & 0.3 & 77.8 & 16.1 \\
& Log Prob & 36.8 & \underline{12.7} & 55.3 & 50.6 & 34.2 & \underline{20.9} & 58.8 & 51.8 & -- & \underline{0.3} & 74.3 & 18.8 & -- & \underline{0.6} & 78.0 & 21.8 \\
& Byte Dist & \underline{73.2} & \textbf{43.7} & \textbf{48.9} & \underline{95.6} & \underline{71.2} & \textbf{58.5} & \textbf{54.2} & \underline{96.3} & -- & \textbf{5.0} & \textbf{72.8} & \underline{69.0} & -- & \textbf{6.2} & \textbf{76.3} & \underline{77.1} \\
& \Decaf ReRanker & \textbf{78.7} & \textbf{43.7} & \underline{51.0} & \textbf{96.2} & \textbf{73.5} & \textbf{58.5} & \underline{54.9} & \textbf{97.4} & -- & \textbf{5.0} & \underline{74.2} & \textbf{71.9} & -- & \textbf{6.2} & \underline{76.8} & \textbf{78.9} \\
\midrule
\multirow{4}{*}{Unstripped}
& None (N=1) & 45.9 & 18.1 & 44.5 & 60.6 & 38.8 & 24.7 & 46.9 & 56.9 & 25.6 & 16.2 & 35.5 & 30.5 & 22.0 & 12.0 & 50.0 & 36.0 \\
& Log Prob & 52.3 & \underline{25.1} & 39.7 & 64.6 & 47.6 & \underline{31.9} & 43.0 & 64.9 & 29.0 & \underline{19.3} & \textbf{31.9} & \underline{33.2} & 25.8 & \underline{14.1} & \underline{48.3} & 40.9 \\
& Byte Dist & \underline{75.7} & \textbf{49.6} & \textbf{36.7} & \underline{93.1} & \underline{74.0} & \textbf{60.9} & \textbf{41.4} & \underline{92.2} & \underline{41.8} & \textbf{28.4} & \underline{32.7} & \textbf{50.6} & \underline{35.8} & \textbf{19.9} & \textbf{48.2} & \underline{61.9} \\
& \Decaf ReRanker & \textbf{79.4} & \textbf{49.6} & \underline{38.4} & \textbf{93.8} & \textbf{76.3} & \textbf{60.9} & \underline{41.8} & \textbf{93.0} & \textbf{44.5} & \textbf{28.4} & 33.2 & \textbf{50.6} & \textbf{38.0} & \textbf{19.9} & 49.0 & \textbf{62.0} \\
\bottomrule
\end{tabular}
}
\end{table*}

%% file: tables/juliet_aggregate.tex
\begin{table}[ht!]                                                                                                                                                                                                                                                                                                                               
  \centering                                                                                                                                                                                                                                                                                                                                  
  \caption{Vulnerability recovery on the Juliet test suite. Ghidra denotes the Ghidra decompiler with our custom exporter. Metrics: F1 (F1 score), P (Precision \%), R (Recall \%), Comp (Compile \%). Best per column in \textbf{bold}.}
  \label{tab:juliet_table}
  \small                                                                                                                                                                                                                                                                                                                                        
  \setlength{\tabcolsep}{4pt}                                                                                                                                                                                                                                                                                                                 
  \begin{tabular}{l c c c c}                                                                                                                                                                                                                                                                                                                    
  \toprule                                                                                                                                                                                                                                                                                                                                      
  Method & F1 & P & R & Comp \\                                                                                          
  \midrule                                                                                                                                                                                                                                                                                    
  Ghidra                          & 1.7              & \textbf{100.0}   & 0.8           & 87.4             \\
  \Decaf-22B (Byte Dist)          & 25.0             & \textbf{100.0}   & 14.3          & \underline{97.3} \\
  \Decaf-22B (Log Prob)           & \underline{29.6} & 91.3             & \textbf{17.6} & 83.2             \\
  \Decaf-22B (\Decaf ReRanker)    & \textbf{29.8}    & \underline{95.5} & \textbf{17.6} & \textbf{98.2}    \\
  \bottomrule                                                                                                                                                                                                                                                                                                                                   
  \end{tabular}
\end{table}


%% file: tables/clang_compile_rerank.tex
\begin{table*}[h!]
\centering
\caption{Reranking results with Clang recompilation. Generated decompilations are recompiled with \texttt{clang -O2} or \texttt{clang -Os} before evaluation. N=32 candidate samples. \textbf{None}: single-sample baseline (N=1). \textbf{Log Prob}: reranking by model log probability. \textbf{Byte Dist}: reranking by bytewise distance to original binary. \textbf{\Decaf ReRanker}: learned reranking model (using stripped reranker for unstripped data). Exact match omitted as recompilation precludes binary-identical outputs.}
\label{tab:clang_reranking}
\resizebox{\textwidth}{!}{
\begin{tabular}{ll ccc ccc ccc ccc}
\toprule
& & \multicolumn{6}{c}{\textbf{Real}} & \multicolumn{6}{c}{\textbf{Synth}} \\
\cmidrule(lr){3-8} \cmidrule(lr){9-14}
& & \multicolumn{3}{c}{-O2} & \multicolumn{3}{c}{-Os} & \multicolumn{3}{c}{-O2} & \multicolumn{3}{c}{-Os} \\
\cmidrule(lr){3-5} \cmidrule(lr){6-8} \cmidrule(lr){9-11} \cmidrule(lr){12-14}
Split & Method & Acc & Edit & Comp & Acc & Edit & Comp & Acc & Edit & Comp & Acc & Edit & Comp \\
\midrule
\multirow{4}{*}{Stripped}
& None (N=1) & 59.1 & 59.2 & 95.4 & 59.1 & 59.2 & 95.4 & -- & 62.9 & 88.9 & -- & 62.9 & 88.9 \\
& Log Prob & \underline{64.2} & \underline{58.3} & 96.0 & \underline{64.2} & \underline{58.3} & 96.0 & -- & \textbf{60.1} & 91.3 & -- & \textbf{60.1} & 91.3 \\
& Byte Dist & 63.7 & \textbf{57.8} & \underline{98.9} & 63.7 & \textbf{57.3} & \underline{99.6} & -- & \underline{62.0} & \underline{96.7} & -- & \underline{61.8} & \underline{96.6} \\
& \Decaf ReRanker & \textbf{70.5} & 58.5 & \textbf{99.8} & \textbf{70.2} & 58.4 & \textbf{99.8} & -- & 62.7 & \textbf{97.1} & -- & 62.9 & \textbf{97.1} \\
\midrule
\multirow{4}{*}{Unstripped}
& None (N=1) & 53.8 & 49.5 & 94.7 & 53.8 & 49.5 & 94.7 & \underline{47.5} & 40.6 & 90.1 & \underline{47.5} & 40.6 & 90.1 \\
& Log Prob & \textbf{66.1} & \textbf{45.6} & 96.8 & \textbf{66.1} & \textbf{45.6} & 96.8 & \textbf{62.5} & \textbf{36.1} & \textbf{95.0} & \textbf{62.5} & \textbf{36.1} & \textbf{95.0} \\
& Byte Dist & \underline{60.6} & \underline{47.9} & \underline{98.1} & \underline{64.2} & \underline{47.4} & \underline{99.4} & 46.0 & \underline{38.5} & 94.4 & 45.5 & \underline{38.3} & 94.5 \\
& \Decaf ReRanker & 60.4 & 49.4 & \textbf{99.7} & 60.4 & 49.6 & \textbf{99.7} & 42.7 & 41.4 & \underline{94.6} & 42.8 & 41.4 & \underline{94.6} \\
\bottomrule
\end{tabular}
}
\end{table*}

%% file: sections/discussion.tex
\section{Discussion and Future Work}

Taking more samples from a neural LLM decompiler increases the potential for both functionally and idiomatically correct decompilations. Given access to the correct compiler configuration, automatic feedback and a trained neural reranker are highly effective at boosting both functional correctness and similarity to the original source code. Even without compiler feedback, Log Probability reranking dramatically improves over single-sample inference, and our reranking methods generalize effectively to other models such as \llmfordecompilemodel.

\noindent\textbf{Performance on Compilers Outside the Training Set}. Our Clang stress-test (\Cref{subsec:stress_test}) reveals mixed results: the neural reranker's performance drops, likely due to distribution shift from unfamiliar Clang-generated assembly patterns. Extending both the generator and reranker to diverse compiler configurations and instruction set architectures is an important direction for future work if the reranker is intended to generalize to these domains. 

\noindent\textbf{Reinforcement Learning and Iterative Refinement}. As described in \Cref{subsec:data}, we sample eight candidates per function during data collection, yielding positive and negative sequences that could be used for offline RL \cite{peng2019advantage, chen2021decisiontransformer}, Direct Preference Optimization \cite{rafailov2023dpo}, or online RL algorithms such as PPO \cite{schulman2017ppo, stiennon2020rlhf_summarization} or GRPO \cite{shao2024deepseekmath_grpo}. In earlier iterations we also explored an iterative refinement model to recursively edit generator outputs, but did not find significant improvements.

%% file: sections/background_related.tex
\section{Related Work}

\subsection{Traditional Decompilation}
\label{sec:rel:trad_decompilation}

Decompilation has been studied for over five decades~\cite{VanEmmerik2001},
but most modern systems trace their lineage to Cifuentes' 1994 dissertation~\cite{Cifuentes1994}.
Cifuentes argued that decompilers should mirror compiler architecture: a
front-end that translates binary or assembly into an intermediate
representation, an analysis layer, and a back-end that emits high-level code.
The roles are reversed compared to a compiler—the decompiler front-end consumes
low-level code and the back-end reconstructs source.  Cifuentes also highlighted
several critical stages (type analysis, control-flow graph recovery, and
control-flow structuring) that remain central to contemporary decompilers.

Recent research has explored many problems in decompilation relevant to
security.  A major area is \emph{control-flow structuring}: converting a
control-flow graph into high-level constructs such as while and if. Schwartz
\etal proposed an iterative algorithm that inserts goto statements where
necessary to preserve semantics~\cite{schwartz:2013}.  Yakdan \etal showed that
some of those gotos can be avoided by duplicating code regions~\cite{dream}.
More recently, Basque \etal observed that certain gotos are artifacts of
compiler transformations and proposed a compiler-aware structuring algorithm
to recognize and reverse those patterns~\cite{sailr}.

Type analysis is another area of research in academic decompilers.
Caballero and Lin survey the area and organize approaches by inferred types,
methods, and evaluation practices~\cite{tiesurvey}.  Early work such as
Mycroft's type-based decompilation frames reconstruction as classic type
inference over recovered program structure~\cite{mycroft:1999}.  TIE uses type
reconstruction theory as well, but explicitly recovers source-level variables
instead of registers and reports a \emph{range} of possible types to cope with
uncertainty~\cite{lee:2011}.  OSPREY
distinguishes itself by applying probabilistic analysis to recover variables
and data structures from stripped binaries~\cite{osprey}.  ReTypd and a closely related work, BinSub, focus on recovering \emph{polymorphic} types~\cite{retypd,binsub}. T-Rex is unique in its explicit shift from recovering a single ground-truth
source type to \emph{type reconstruction}: it uses structural types to capture
behavior and introduces a new evaluation framing and metric aligned with
reverse-engineering needs~\cite{trex}.

In practice, security practitioners often rely on industrial decompilers such as
Ghidra~\cite{ghidra}, Hex-Rays~\cite{hexrays}, and Binary Ninja~\cite{binja}.
While these tools differ in user experience and internal representations, they
share the same overall objective as academic systems: recover readable,
high-level code from low-level machine instructions by combining control-flow
recovery, dataflow analysis, and type recovery.  Ironically, although these
tools are widely used in security practice, their internal algorithms and
evaluation methodologies are not well documented in the academic literature,
with a few exceptions~\cite{guilfanov:2008}.

\subsection{Neural Decompilation}

Neural decompilation has recently emerged as a promising alternative to
traditional program analysis-based approaches.  One of the primary challenges of
decompilation is the \emph{one-to-many} nature of the problem: many different
source programs can compile to the same binary.  Neural models can naturally
capture this uncertainty by learning a distribution over possible source
programs given a binary.  This makes them particularly well-suited to recovering
details that are lost during compilation, such as identifier names, as well as
deciding between typing decisions that can result in identical code (e.g., a
struct with two ints vs. an array of two ints). 

\subsubsection{Abstractions}

Most early work in neural decompilation focused on recovering specific pieces of
information that are lost during the compilation process.  Arguably one of the
most painful omissions of traditional decompilers is the loss of meaningful
identifier names; unsurprisingly, several works have applied neural models of
various architectures to recover such
names~\cite{lacomis2019dire,chen2022dirty,varbert,gennm}.  

Despite much research in type recovery in traditional decompilation
(see~\cref{sec:rel:trad_decompilation}), mainstream decompilers continue to
struggle to recover helpful type information from stripped binaries~\cite{trex}.
This can also be seen in our working example in~\cref{fig:working}.  Recent work
has applied neural models to improve the type recovery
process~\cite{chen2022dirty,typeforge,tygr,resym,dramko2025idioms}.  Such work
has two notable advantages.  First, similar to identifier name recovery, neural
models can not only recover lost type information but can also predict
meaningful type and field \emph{names}, which can have a significant impact on
code readability.  Second, neural models can better handle the inherent
uncertainty in type recovery by learning distributions over possible types than
most traditional approaches which are not statistical (with the notable
exception of OSPREY~\cite{osprey}).  

\subsubsection{Decompilation}

Some of the earliest work in neural decompilation focused on end-to-end decompilation.  These systems attempted to directly translate binaries or assembly code into high-level source code using neural sequence-to-sequence models~\cite{katz2018,beyondthec}.

More recently, with the introduction of large language models (LLMs) and their
demonstrated ability to generate code, there has been a surge of interest in
using LLMs for decompilation tasks.  These models leverage vast amounts of
training data and sophisticated architectures to produce high-quality
decompilations~\cite{armengol2024slade,tan2024llm4decompile,dramko2025idioms}.
Most recent work~\cite{tan2024llm4decompile,dramko2025idioms}, including ours,
trains a model to start from the output of a traditional decompiler (versus
assembly code~\cite{armengol2024slade}).  The benefits of this hybrid approach
are multifold.  First, it leverages the long history of research and engineering
that has gone into traditional decompilers, rather than starting from scratch.
Second, decompiler output is generally much shorter than raw assembly or binary
code, making the learning problem easier for neural models.  Finally, as we show
in our experiments in \Cref{tab:results}, using a custom exporter, Ghidra's decompiled code
is difficult to read but is functionally accurate, making it an excellent
starting point for neural refinement.

Our \DecafLLM model is similar in spirit to these recent LLM-based decompilation
systems but differs in a few critical ways.  First and foremost, our model is
trained to produce compilable code, which we believe is a critical requirement
for practical use of decompilation tools.
LLM4Decompile~\cite{tan2024llm4decompile} often references types without
defining them; Idioms~\cite{dramko2025idioms} improves this by
recovering type definitions separately, but their final decompilations may still
often fail to compile.  Last, our model is designed to work
in tandem with a neural reranker, which provides another opportunity for our
approach to yield a more accurate decompilation.

\subsection{Perfect Decompilation}

While most pre-neural decompilation work emphasizes \emph{program analysis},
another line of work targets \emph{perfect} (byte-equivalent) decompilation,
where recompiling the recovered source yields a binary identical to the target.
Decomperson~\cite{decomperson} ran a large-scale competition to study how
experts iteratively refine decompilations toward this goal, and the video-game
reverse-engineering community has long pursued perfect decompilation to
reconstruct lost source code for classic games~\cite{decompme}. Schulte \etal
used genetic programming to evolve code snippets and \emph{automate} the search
for perfect decompilations~\cite{schulte2018evolving}.  Our approach is
conceptually similar to this work but replaces genetic programming with
temperature-based sampling of neural models.  We also show that our neural
reranker outperforms the simpler edit-distance-based strategies.

\noindent\textbf{Compiler Provenance}. \label{sec:rel:provenance}
One of the underlying assumptions of perfect decompilation is that the original
compilation process can be repeated.  This of course requires the original
compiler and flags to be known.  Fortunately, there is prior work on
recovering information about the compiler used to build a binary and its
flags~\cite{rosenblum2010compilerprovenance,du2023compilerprovenance}. However, use of a compiler that is unavailable for any
reason could still pose challenges for perfect decompilation.

\subsection{AI and Machine Learning}

Our approach—sampling many candidates and selecting the best via a learned reranker, commonly referred to a ``verifier" in the machine learning literature, is an instance of test-time compute scaling \cite{snell2024scaling}. An early instance of generate-then-verify paradigm was introduced for math reasoning \cite{cobbe2021training} and also found success with AlphaCode in addressing competitive programmings~\cite{li2022competition}, by generating a very large amount of programs and filtered by execution. 

%% file: sections/threats_and_conclusion.tex
\section{Threats to Validity}
\label{sec:threats-to-validity}

In our work, we choose to use \Exebench, because it has been used in the neural decompilation literature \cite{tan2024llm4decompile, armengol2024slade, dramko2025idioms}, it is mined from real world open-source software projects, and it ships with test cases. Although the dataset was intended for purposes such as decompilation, the test cases in \Exebench may not be comprehensive: a lack of test case coverage may admit false positives, thereby increasing reported functional correctness scores. Additionally, while the functions in \Exebench are mined from real world software projects, the implicit filtering and processing necessary to collect and prepare the data may not reflect the complexity of source code encountered in real-world reverse engineering scenarios. Because adversarial code such as malware is not as publicly available and likely under-represented in \Exebench we do not claim our model will generalize to malware patterns that it has not been trained on. We chose the Juliet test suite because it is a well-established benchmark; however, given that many test cases may be synthetic, performance on vulnerability recovery may not generalize to more complex code bases. 

Because of the large volume of already-existing experiments, we did not collect or evaluate on functions compiled for alternate architectures like ARM or RISC-V. Furthermore, we do not employ more advanced techniques to make reversing more challenging such as obfuscation.

\section{Conclusion}

We presented \Decaf, a system that improves neural decompilation by generating multiple candidate decompilations and automatically selecting the most promising one using compiler feedback and a trained neural reranker. Our results demonstrate that taking more samples from a neural decompiler dramatically increases the chance of producing both functionally and idiomatically correct code. We show that provided access to the reference compiler, that automatic feedback, whether through bytewise distance or neural reranking, is highly effective at identifying the best candidate. Importantly, our reranking methods generalize to other models such as \llmfordecompilemodel, and even Log Probability reranking yields substantial gains over single-sample inference. We also demonstrate that our model can be useful in a practically-framed downstream task such as vulnerability recovery. We plan to release our models, data, and evaluation infrastructure to support the community and future work in neural decompilation.

\clearpage